\input{aipcheck}

\documentclass[
    ,final            
  ]
  {aipproc}

\layoutstyle{8x11double}

\begin{document}

\title{Flavor structure with multi moduli in 5D SUGRA\footnote{Talks 
 given by Y.S. at PASCOS'08 (Perimeter Institute, Canada, June 2-6, 2008) 
 and at SUSY'08 (Seoul, Korea, June 16-21, 2008).}}


\author{Hiroyuki Abe}{
  address={Yukawa Institute for Theoretical Physics, Kyoto 606-8502, Japan}
}

\author{Yutaka Sakamura}{
  address={RIKEN, Wako, Saitama 351-0198, Japan}
}


\begin{abstract}
We investigate 5-dimensional supergravity on $S^1/Z_2$ 
with a physical $Z_2$-odd vector multiplet, 
which yields an additional modulus other than the radion. 
We find additional terms in the 4-dimensional effective theory 
that are peculiar to the multi moduli case. 
Such terms can make the soft masses are non-tachyonic and 
almost flavor-universal at tree-level, in contrast to 
the single modulus case. 
This provides a new possibility to solve the SUSY flavor problem.  
\end{abstract}

\maketitle


\section{Introduction}
One of the simplest setup for the extra-dimensional model with 
supersymmetry (SUSY) is five-dimensional (5D) supergravity compactified 
on $S^1/Z_2$, 
which is extensively studied in a large number of papers. 
Most of such works assume or consider a situation that 
the radius of the extra dimension~$r$ is determined by 
the vacuum expectation value (VEV) of a single chiral multiplet~$T$ 
(the radion multiplet), that is, $\pi r={\rm Re}\,\langle T \rangle$. 
In general, however, there are also cases where $r$ is determined by 
VEVs of more than one chiral multiplets, such as 
\begin{equation}
 \pi r = {\cal F}({\rm Re}\,\langle T^1\rangle, 
 {\rm Re}\,\langle T^2\rangle, \cdots), \label{r_multi_mod}
\end{equation}
where ${\cal F}$ is some function, and $T^I$ ($I=1,2,\cdots,n$) are 
chiral multiplets which we call the moduli in this talk. 

The single modulus case ($n=1$) corresponds to the case that 
there are no physical 5D vector multiplets whose scalar components have 
zero-modes. 
The radion mode comes from the zero-mode of 
the extra-dimensional component of the f\"{u}nfbein~$e_y^{\;\;4}$. 
When some scalar components of physical 5D vector multiplets have zero-modes, 
the radion mixes with them to form chiral multiplets in the 
four-dimensional (4D) effective theory. 
Thus $n$ moduli consist of zero-modes for $e_y^{\;\;4}$ and 
$(n-1)$ scalar components of the 5D vector multiplets. 
In this case the orbifold radius~$r$ is given by a combination of VEVs of 
the $n$ moduli as shown (\ref{r_multi_mod}). 

In the multi moduli case ($n\geq 2$), the low-energy physics can be 
changed from that in the single modulus case 
due to the existence of the physical 5D vector multiplets. 
In this talk we investigate 4D effective theory of 5D supergravity 
in the multi moduli case and 
see the difference from that in the single modulus case. 
Especially we focus on the flavor structure 
of the soft SUSY breaking masses~\cite{AS:2008}.

\section{Setup}
The off-shell formulation is useful 
to describe the 5D supergravity action. 
Here we adopt the conformal supergravity formulation developed by 
Ref.~\cite{KO}. 
In this formulation, the $(n-1)$ physical 5D vector multiplets 
are expressed as $n$ off-shell vector multiplets 
by adding unphysical degrees of freedom. 
One combination of the $n$ vector components is identified with 
the graviphoton, which belongs to the supergravity multiplet 
in the on-shell description. 

As the simplest case, we consider the two moduli case. 
Namely we introduce two vector multiplets~${\cal V}^1=(V^1,\Sigma^1)$, 
${\cal V}^2=(V^2,\Sigma^2)$ which include the moduli as zero-modes. 
We also introduce a hypermultiplet~$(X,X^c)$ 
which is relevant to the SUSY breaking 
besides the matter hypermultiplets~$(Q_i,Q_i^c)$. 
The index~$i$ run over quarks, leptons and Higgses. 
We use $N=1$ superfield notation to express each 5D multiplet. 
$V^1$, $V^2$ are $N=1$ vector multiplets while the others are 
chiral multiplets. 
Among them, $\Sigma^1$, $\Sigma^2$, $X$ and $Q_i$ are even 
under the $Z_2$-parity around the orbifold boundaries, 
while the others are $Z_2$-odd. 
Thus the former have zero-modes~$T^1$, $T^2$, $X_0$ and $Q_{0i}$, 
respectively.\footnote{
The standard model gauge multiplets are contained in 
5D vector multiplets whose vector components are $Z_2$-even. 
For simplicity, we omit them in this talk because they are irrelevant to 
the tree-level soft masses. }
 
In 5D supergravity, every mass scale in the bulk Lagrangian is introduced 
as a gauge coupling constant (in the unit of the 5D Planck mass) 
for some vector multiplet whose lowest component is $Z_2$-even. 
Thus we asume that $(Q_i,Q_i^c)$ and $(X,X^c)$ are charged 
for ${\cal V}^1$ with the gauge couplings~$c_i$ and $c_X$ 
in order to introduce the bulk mass parameters 
for the hypermultiplets.\footnote{
Of course we can also gauge the hypermultiplets by ${\cal V}^2$. }
The localization of the wave functions for $Q_{0i}$ and $X_0$ in the bulk 
is controlled by these bulk mass parameters~$c_i$ and $c_X$ 
and the hierarchical structure of the Yukawa couplings can be realized 
just in the similar way to Ref.~\cite{Arkani-Schmaltz}.
Besides these gauging, 
5D supergravity is characterized by a cubic polynomial called 
the norm function:
\begin{equation}
 {\cal N}({\cal V}) \equiv C_{IJK}{\cal V}^I{\cal V}^J{\cal V}^K. 
\end{equation}
A real constant tensor~$C_{IJK}$ is completely symmetric 
for the indices~$I=1,2$.

\subsection{4D effective theory}
The multi moduli case has not been studied very much 
just because of a technical reason. 
In that case the derivation of the 4D effective theory 
becomes much more complicated than that in the single modulus case. 
In our previous work~\cite{AS:2006}, we developed a systematic method 
to derive the 4D effective theory for general 5D supergravity model, 
which is based on an $N=1$ superspace description of 5D conformal 
supergravity~\cite{AS:2004} and developed 
in subsequent works~\cite{AS2}. 
The procedure is as follows. 
We start from the $N=1$ off-shell description of 5D action. 
After some gauge transformation, we drop kinetic terms for $Z_2$-odd 
multiplets which are negligible at low energy. 
Then these multiplets play a role of Lagrange multiplier and 
their equations of motion extract zero-modes from the $Z_2$-even multiplets. 
After these steps, we obtain the following K\"{a}hler potential 
in the 4D effective theory. 
\begin{eqnarray}
 \Omega \!\!\!&\equiv\!\!\!& -3e^{-K_{\rm eff}/3} \nonumber\\
 \!\!\!&=\!\!\!& {\cal N}^{1/3}\left\{-3
 +\sum_i Z_i|Q_{0i}|^2 
 +\sum_i\tilde{\Omega}_{iX}|Q_{0i}|^2|X_0|^2\right\} \nonumber\\
 &&+{\cal O}(|Q_0|^4,|X_0|^4), 
\end{eqnarray}
where 
\begin{eqnarray}
%
 Z_i \!\!\!&=\!\!\!& 
 \frac{1-e^{-2c_i{\rm Re}\,T^1}}{c_i{\rm Re}\,T^1} 
\end{eqnarray}
are coefficients of $|Q_{0i}|^2$ in the effective K\"{a}hler 
potential, i.e., $K_{\rm eff}=-3{\cal N}^{1/3}
+\sum_iZ_i|Q_{0i}|^2+\cdots$. 
The arguments of ${\cal N}$ and $\tilde{\Omega}_{iX}$ are 
$({\rm Re}\,T^1,{\rm Re}\,T^2)$. 
The essential difference between the single and multi moduli cases 
appears in $\tilde{\Omega}_{iX}$.

\section{Soft SUSY breaking masses}
In this talk, we assume that $\langle X_0\rangle\ll 1$ in the unit of 
the 5D Planck mass and the F-term of $X_0$ is 
the dominant source of SUSY breaking. 
In Ref.~\cite{AS:2008}, we provide a specific exalmple that 
realizes such a situation explicitly. 
Under this assumption, the soft masses are expressed as 
\begin{equation}
 m_{{\rm soft}\,i}^2 \simeq -|F^X|^2\frac{\tilde{\Omega}_{iX}}{Z_i}. 
 \label{m_soft}
\end{equation}

Let us first review the single modulus case. 
In this case, we obtain 
\begin{eqnarray}
 {\cal N} \!\!\!&=\!\!\!& ({\rm Re}\,T^1)^3, \nonumber\\
 \tilde{\Omega}_{iX} \!\!\!&=\!\!\!& \frac{1-e^{-2(c_i+c_X){\rm Re}\,T^1}}
 {3(c_i+c_X)}.  \label{single_Omg}
\end{eqnarray}
We can see from (\ref{m_soft}) that the soft masses are tachyonic 
irrespective of the values of the bulk mass parameters~$c_i$ and $c_X$. 
Thus we have to choose $c_i$ and $c_X$ such that 
$e^{-2c_i{\rm Re}\,T^1}\gg 1$ and $e^{-2c_X{\rm Re}\,T^1}\ll 1$ 
so that the tree-level contribution~(\ref{m_soft}) is exponentially 
suppressed and the quantum effects dominate. 
Such quantum effects can save 
the tachyonic masses at tree level. 
From the 5D point of view, the condition~$e^{-2c_i{\rm Re}\,T^1}\gg 1$ 
and $e^{-2c_X{\rm Re}\,T^1}\ll 1$ means that 
the wave functions for $Q_{0i}$ and $X_0$ 
are localized around the opposite boundaries of $S^1/Z_2$. 
Namely the SUSY breaking sector is geometrically sequestered from 
our visible sector. 
This is the situation usually considered in most of the works on 
5D supergravity models. 

In the multi moduli case, the situation is quite different. 
The expression of $\tilde{\Omega}_{iX}$ becomes more complicated. 
\begin{eqnarray}
 \tilde{\Omega}_{iX} \!\!\!&=\!\!\!& \frac{{\cal N}{\cal N}_{11}}
 {3{\cal N}{\cal N}_{11}-2{\cal N}_1^2}
 \frac{1-e^{-2(c_i+c_X){\rm Re}\,T^1}}{(c_i+c_X){\rm Re}\,T^1}
 \nonumber\\
 &&\hspace{-10mm}
 -\frac{{\cal N}{\cal N}_1^2}{3{\cal N}{\cal N}_{11}-2{\cal N}_1^2}
 \frac{(1-e^{-2c_i{\rm Re}\,T^1})(1-e^{-2c_X{\rm Re}\,T^1})}
 {3c_ic_X({\rm Re}\,T^1)^2}, \nonumber\\ \label{multi_Omg}
\end{eqnarray}
where ${\cal N}_1(X)\equiv\partial{\cal N}/\partial X^1$ and 
${\cal N}_{11}(X)\equiv\partial^2{\cal N}/(\partial X^1)^2$. 
The first term has a similar structure to the single modulus case, 
but now it can be negative if $3{\cal N}{\cal N}_{11}
-2{\cal N}_1^2<0$.\footnote{
Note that we have to choose the norm function 
such that ${\cal N}_{11}>0$ in order to obtain 
the correct signs of kinetic terms 
for the scalar components of the 5D vector multiplets. }
Even if it is positive, $\tilde{\Omega}_{iX}$ becomes negative 
when the second term dominates over the first term. 
Especially when $e^{-2c_i{\rm Re}\,T^1}\gg 1$ 
and $e^{-2c_X{\rm Re}\,T^1}\ll 1$, 
the second term of (\ref{multi_Omg}) dominates and 
furthermore the $c_i$-dependence is cancelled in (\ref{m_soft}). 
Namely we obtain non-tachyonic and flavor universal soft massses. 
As mentioned above, this situation corresponds to the geometrical 
separation of the SUSY breaking source~$F^X$ from our visible world. 

In contrast to the single modulus case, the tree-level contribution 
to the soft masses~(\ref{m_soft}) remains finite 
even when $Q_{0i}$ and $X_0$ are localized around the opposite 
boundaries. 
This fact indicates the existence of some heavy modes that couple to both 
$Q_{0i}$ and $X_0$. 
After integrating them out, additional contribution to 
the contact terms~$|Q_{0i}|^2|X_0|^2$ in the K\"{a}hler potential 
appears. 
This corresponds to the difference between (\ref{single_Omg}) 
and (\ref{multi_Omg}). 
Such heavy modes are the Kaluza-Klein modes for the $Z_2$-odd vector 
multiplets~$(V^1,V^2)$. 
Notice that one combination of ${\cal V}^1$ and ${\cal V}^2$ 
is identified with the graviphoton multiplet, which does not contribute to
the contact terms~$|Q_{0i}|^2|X_0|^2$. 
Thus the additional contribution mentioned above 
exists only in the multi moduli case.

\section{Summary}
We study 4D effective theory of 5D supergravity on $S^1/Z_2$ 
in the case that the orbifold radius is determined by VEVs 
of more than one chiral multiplets. 
In such a multi moduli case, 
the flavor structure of the soft SUSY breaking parameters 
is quite different from the single modulus case that is usually considered. 
The additional terms appear in the effective K\"{a}hler potential 
and it can make the soft masses non-tachyonic and almost flavor-universal 
in contrast to the single modulus case. 
This fact provides a new possibility to solve the SUSY flavor problem. 

In this talk, we do not discuss the other soft SUSY breaking parameters, 
such as the A-terms and the gaugino masses. 
In order to study them, 
we need to consider a specific model that stabilizes the moduli 
and breaks SUSY. 
We provide such a model in Ref.~\cite{AS:2008} and discuss 
the possibility to solve the SUSY flavor problem. 
Detailed phenomenological analysis based on this possibility 
is in progress. 




\begin{theacknowledgments}
This work was supported in part by the Japan Society for the Promotion 
of Science for Young Scientists No.182496 (H.A.), and 
by the Special Postdoctoral Researchers Program at RIKEN (Y.S.). 
\end{theacknowledgments}




\bibliography{sample}

\begin{thebibliography}{9}

\bibitem{AS:2008} H.~Abe and Y.~Sakamura, arXiv:0807.3725. 

\bibitem{KO} T.~Kugo and K.~Ohashi, 
 \emph{Prog.~Theor.~Phys.}~\textbf{105} 323 (2001); \textbf{106} 671 (2001);
 \textbf{108} 203 (2002). 

\bibitem{AS:2006} H.~Abe and Y.~Sakamura, \emph{Phys.~Rev.}~\textbf{D75}
 025018 (2007). 

\bibitem{Arkani-Schmaltz} N.~Arkani-Hamed and M.~Schmaltz, 
 \emph{Phys.~Rev.}~\textbf{D61} 033005 (2000); D.E.~Kaplan and T.M.P.~Tait, 
 \emph{JHEP}~\textbf{0006} 020 (2000). 

\bibitem{AS:2004} F.~Paccetti Correia, M.G.~Schmidt and 
 Z.~Tavartkiladze, \emph{Nucl.~Phys.}~\textbf{B709} 141 (2005);
H.~Abe and Y.~Sakamura, \emph{JHEP}~\textbf{0410} 013 (2004). 

\bibitem{AS2} H.~Abe and Y.~Sakamura, \emph{Phys.~Rev.}~\textbf{D71} 
 105010 (2005);\textbf{D73} 125013 (2006). 
\end{thebibliography}



\end{document}